\documentclass[conference]{IEEEtran}

\usepackage{cite,graphicx,amsmath,psfrag,bm}
\usepackage[usenames,dvipsnames]{xcolor}
\usepackage{mathabx}
\usepackage{subfig}
\usepackage{cancel}
\usepackage{epstopdf}
\usepackage{pstool}

\newcommand{\ve}[1]{\boldsymbol{#1}}
\newcommand{\te}[1]{\overline{\overline{#1}}}

\makeatletter\def\CT{\def\@captype{figure}}\makeatother

\begin{document}

\title{Metasurface Spatial Processor \\ for Electromagnetic Remote Control}

\author{\IEEEauthorblockN{Karim Achouri\IEEEauthorrefmark{1}, Guillaume Lavigne\IEEEauthorrefmark{1}, Mohamed A. Salem\IEEEauthorrefmark{1} and
Christophe Caloz\IEEEauthorrefmark{1}\IEEEauthorrefmark{2}}
\IEEEauthorblockA{\begin{minipage}{6.5cm}
\centering\linespread{1}
\IEEEauthorrefmark{1}Dept. of Electrical Engineering \\
Polytechnique Montr\'{e}al \\
Montr\'{e}al, QC H2T 1J3, Canada
\end{minipage}
\begin{minipage}{6.5cm}
\centering
\IEEEauthorrefmark{2}Electrical and Computer Engineering Dept. \\
King Abdulaziz University \\
POB 80204, Jeddah 21589, Saudi Arabia
\end{minipage}}}

\maketitle

\begin{abstract}
We introduce the concept of metasurface spatial processor, whose transmission is remotely and coherently controlled by the superposition of an incident wave and a control wave through the metasurface. The conceptual operation of this device is analogous to both that of a transistor and a Mach-Zehnder interferometer, while offering much more diversity in terms of electromagnetic transformations. We demonstrate two metasurfaces, that perform the operation of electromagnetic switching and amplification.
\end{abstract}

\IEEEpeerreviewmaketitle

%\tableofcontents

\section{Introduction}

Over the past decade, metasurfaces~\cite{PhysRevLett.93.197401,Holloway2009,holloway2012overview,Pfeiffer2013a,yu2014flat} have emerged as an outgrowth of volume metamaterials~\cite{caloz2005electromagnetic,engheta2006metamaterials,capolino2009theory}, in a similar fashion as two-dimensional photonic crystals emerged from three-dimensional photonic crystals about a decade earlier~\cite{joannopoulos2011photonic}, because of their benefits of reduced loss and easier fabrication. However, in contrast to what happened in photonic crystals, dimensional reduction in metasurfaces has brought about enriched rather than reduced properties and functionalities.

Metasurfaces are two-dimensional structures composed of generally nonuniform periodic or quasi-periodic arrangements of scattering particles engineered to transform the scattered field according to specifications when illuminated by a specified incident field. Their structure is fundamentally similar to that of frequency selective surfaces~\cite{MunkFSS}, but they exhibit many more processing capabilities (e.g. generalized refraction, birefringence, orbital angular momentum multiplexing, spatial dispersion processing, etc.) by exploiting a larger number of degrees of freedom, in particular nonuniformity and various additional features such as nonreciprocity and bianisotropy. Several synthesis techniques have been recently developed to synthesize such metasurfaces~\cite{Salem2013c,achouri2014general,PhysRevApplied.2.044011,6477089,6905746}.

We present here the concept of a metasurface spatial processor, that may be seen as a functional extension of both the Mach-Zehnder interferometer~\cite{saleh2007fundamentals} and the transistor~\cite{sze2006physics}. As these devices, this metasurface spatial processor can perform switching and amplifying operations under the application of an external control signal. However, it additionally provides the capabilities of performing these operations remotely, using an electromagnetic control wave, and of providing the whole range of aforementioned metasurface spatial wave transformations.

The remote nature of the control in this metasurface device contrasts with the local control of previously reported tunable metasurfaces based on electrically tuned scattering particles~\cite{Burokur2,PhysRevB.86.195408,6349393,zhu2013active}, while its Mach-Zehnder and transistor functionalities are totally distinct from those of previously reported optically controlled nonlinear metasurfaces~\cite{xie2013spatial,harris2008electromagnetically}. The metasurface spatial processor modulates the incident wave via a coherently superimposed control wave, as done in~\cite{Shi:14} where the addition of coherent wave on the transmit side of the metasurface is used to suppress undesired reflected or transmitted waves.

The paper is organized as follows. First, we introduce the concept of the metasurface spatial processor. Next, we discuss its implementation of the switching and amplifying operations. Then, we addresses the mathematical synthesis as well as the physical realization of the metasurface. Finally, we present full-wave simulations and experimental results that confirm the expected theoretical results. The time convention $e^{-i\omega t}$ is assumed throughout the paper.

\section{Metasurface Spatial Processor Concept}

\begin{figure}[h!]
\centering
\subfloat[]{
\includegraphics[width=0.4\linewidth]{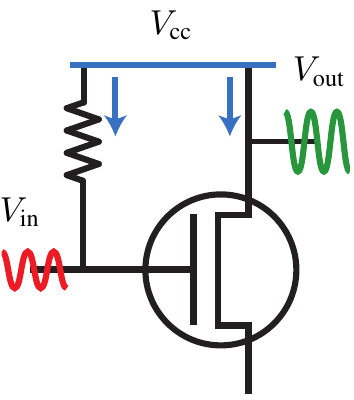}
%\psfragfig*[width=0.4\columnwidth]{trans2}{
%\psfrag{a}[c][c][1]{$V_\text{in}$}
%\psfrag{b}[c][c][1]{$V_\text{out}$}
%\psfrag{c}[c][c][1]{$V_\text{cc}$}}
\label{fig:param_prob1}}
\qquad
\subfloat[]{\raisebox{3mm}{
\includegraphics[width=0.4\linewidth]{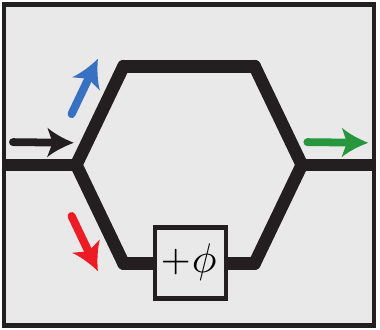}
%\psfragfig*[width=0.43\columnwidth]{MZ}{
%\psfrag{a}[c][c][1]{$V_\text{in}$}
%\psfrag{b}[c][c][1]{$V_\text{out}$}
%\psfrag{c}[c][c][1]{$+V_\text{cc}$}
%\psfrag{d}[c][c][1.2]{$+\phi$}
%\psfrag{e}[c][c][1]{$+V_\text{cc}$}}
}
\label{fig:param_prob2}}
\\
\subfloat[]{\includegraphics[width=0.4\linewidth]{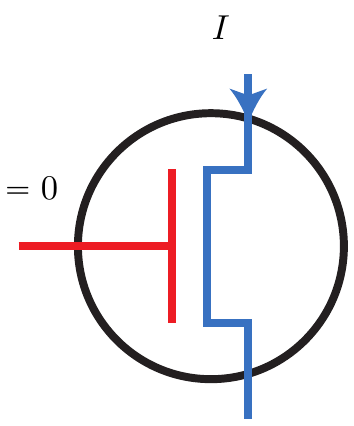}
\label{fig:param_prob3}}
\qquad
\subfloat[]{\includegraphics[width=0.4\linewidth]{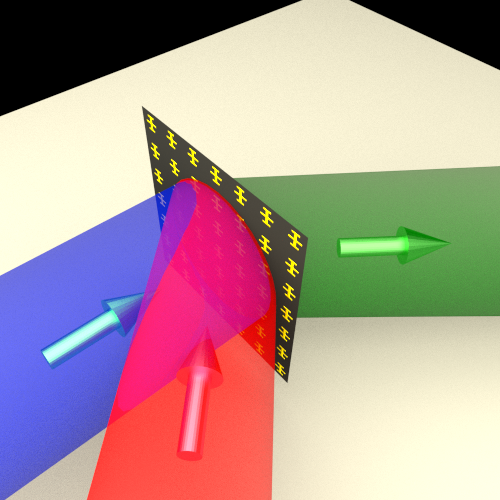}
\label{fig:param_prob4}}
\caption{Metasurface spatial processor as a conceptual functional extension of both the Mach-Zehnder interferometer and the transistor. (a)~Transistor. (b)~Mach-Zehnder interferometer. (c)~Unmodulated metasurface spatial processor transforming an input (incident) wave into an arbitrary output (transmitted and/or reflected) wave. d)~Metasurface spatial processor in the modulated regime, where the output wave is modified by a remote control wave interfering on the metasurface with the input wave.}\label{fig:param_prob}
\end{figure}

The metasurface spatial processor concept is represented in Fig.~\ref{fig:param_prob} as an extension of both the Mach-Zehnder interferometer and the transistor. We suggest here that a metasurface, with its rich field transformation capabilities, represents an ideal opportunity to extend the Mach-Zehnder and transistor devices for complementary and additional functionalities.

The conceptual operation of a conventional transistor is represented in Fig.~\ref{fig:param_prob1}, while that of the metasurface spatial processor is represented in Figs.~\ref{fig:param_prob3} and~\ref{fig:param_prob4}. In the latter case, the incident wave, that illuminates the metasurface, plays the role of the DC bias (Fig.~\ref{fig:param_prob3}) of the transistor. The incident wave opens an ``electromagnetic channel,'' and this channel can be modulated by the application of a control wave (Fig.~\ref{fig:param_prob4}) that modifies the wave transmitted by the metasurface, in a similar fashion as the DC bias of a transistor opens its semiconductor channel that may be next modulated by the application of a dynamic voltage (Fig.~\ref{fig:param_prob1}).

The Mach-Zehnder interferometer shown in Fig.~\ref{fig:param_prob2} is an integrated version of the original Mach-Zehnder that is composed of beam splitters and mirrors. This device may be used as an optical modulator as well as an optical switch. Its operation may be summarized as follows. First, the input wave is split into two beams, a reference beam and a modulation beam whose phase shift ($\phi$) may be controlled. Then, the two beams are recombined coherently to form the output wave as a modulated version of the input wave. The metasurface spatial processor operates in a similar fashion. In addition, as previously mentioned, the coherent superposition of the incident wave and the control wave on the metasurface allows for a sophisticated spatial control scattered waves with a great diversity of possible transformations. For instance, the application of the control wave may be used to deflects the transmitted wave, as shown in Figs.~\ref{fig:param_prob3} and~\ref{fig:param_prob4}.

Note that the operation example in Figs.~\ref{fig:param_prob3} and~\ref{fig:param_prob4} only represents a particular spatial processing operation. In general, the superposition of the incident and control waves can manipulate the amplitude, the phase, the polarization,  and the direction of refraction of the reflected and transmitted waves in an almost arbitrary fashion~\cite{achouri2014general}. The metasurface may also include nonlinear and time-varying elements for transforming the temporal spectrum of the input wave.

\section{Switch-Amplifier Illustration}

To illustrate the metasurface spatial processor concept, we will present the case of a switch-amplifier processor. In this device, the switching operation is achieved by destructive interference between the incident wave $(E_\text{i})$ and the control wave $(E_\text{c})$ while linear amplification is obtained by tuning the phase of the control wave. In this scenario, assuming that the waves propagate along the $z$-direction with propagation constant $k$, the amplitude of the transmitted wave $(E_\text{t})$ changes according to
\begin{equation}
E_\text{t} = E_\text{i}^+ + E_\text{c}^+ = e^{ikz} + e^{ikz + i\phi} = 2e^{ikz+i\phi/2}\cos{\left (\frac{\phi}{2} \right )},
\end{equation}
where $E_\text{i}^+$ and $E_\text{c}^+$ denote the incident and the control waves, respectively, in the transmit side of the system, and where $\phi$ is the phase difference between $E_\text{i}^+$ and $E_\text{c}^+$. The amplitude of the transmitted wave therefore spans the range between $0$ and $2$, which, in the latter limit, corresponds to a gain of $2$ compared to the amplitude of the transmitted wave when only $E_\text{i}$ is illuminating the metasurface.

Because the incident wave and the control wave may impinge on the metasurface under different angles, the metasurface must be able to transform the two waves independently. Therefore, these waves must have mutually orthogonal polarizations on the incident side of the metasurface. However, since they must interfere on the transmit side, the polarization of the control wave must be rotated to match the polarization of the incident wave. In such circumstances, the metasurface differently affects the $x$ and $y$ polarizations (birefringence) and rotates one polarization without affecting the other one (anisotropic chirality).

\begin{figure}[h!]
\begin{center}
\subfloat[]{\label{Fig:TMconcept1}
\includegraphics[width=0.5\columnwidth]{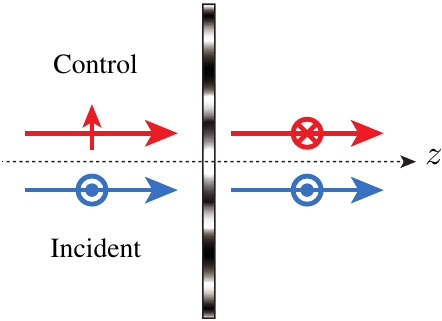}
%\psfragfig*[width=0.5\columnwidth]{TM1}{
%\psfrag{a}[c][c][1]{$z$}
%\psfrag{b}[c][c][0.8]{Incident}
%\psfrag{c}[c][c][0.8]{Control}}
}
\subfloat[]{\label{Fig:TMconcept2}
\includegraphics[width=0.5\columnwidth]{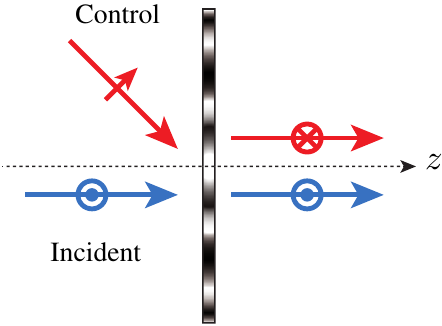}
%\psfragfig*[width=0.5\columnwidth]{TM2}{
%\psfrag{a}[c][c][1]{$z$}
%\psfrag{b}[c][c][0.8]{Incident}
%\psfrag{c}[c][c][0.8]{Control}}
}
\caption{Representations of two illustrative switch-amplifier metasurface spatial processors. In (a), the incident and control waves are both normally incident on the metasurface. The polarization of the control wave is rotated by 90$^\circ$ to match the polarization of the incident wave. A $\pi$-phase shift is imposed between the two transmitted waves so that they cancel out by destructive interference. In (b), the control wave is obliquely impinging on the metasurface, to avoid the unpractical collocation of the incident and control wave sources.}\label{Fig:TMconcept}
\end{center}
\end{figure}

In order to realize this operation, two metasurface configuration are considered. The first configuration, shown in Fig.~\ref{Fig:TMconcept1}, represents the simplest case, where the incident wave and the control wave are both normally incident on the metasurface. As may be seen in the figure, on the left-hand side of the metasurface, the control wave is p-polarized while the incident wave is s-polarized. The metasurface is designed to pass both waves with the same transmission coefficient to ensure complete power extinction or maximal amplification. On the right-hand side, the polarization of the control wave is rotated so that both transmitted waves are s-polarized to interfere. In this scenario, the phase difference between $E_\text{i}^+$ and $E_\text{c}^+$ is such that the control wave suppresses transmission by destructive interference $(\phi = \pi)$. The second configuration, shown in Fig.~\ref{Fig:TMconcept2}, performs the same operation but for the case of an obliquely impinging control wave, that spatially separates the incident and control wave sources, as may be required in practical systems.

\begin{figure}[h!]
\begin{center}
\includegraphics[width=0.7\columnwidth]{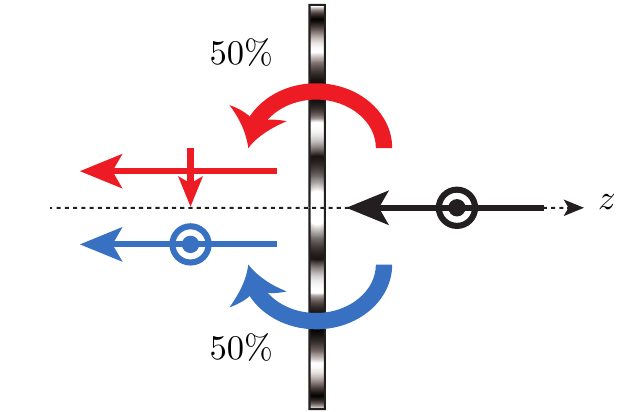}
%\psfragfig*[width=0.7\columnwidth]{TM3}{
%\psfrag{a}[c][c][1]{$z$}
%\psfrag{b}[c][c][1]{$50\%$}}
\caption{Normally incident s-polarized wave impinging on the metasurface from the right (negative $z$-direction) in Fig.~\ref{Fig:TMconcept1}. The incident wave is transformed, by design, into two mutually orthogonal waves with equal amplitude.}\label{Fig:TMconceptReci}
\end{center}
\end{figure}

An important feature of the processor in Fig.~\ref{Fig:TMconcept} is that its efficiency is inherently limited by reciprocity. This may be understood by considering a normally incident s-polarized wave impinging on the metasurface from the right in Fig.~\ref{Fig:TMconcept1}, as shown in Fig.~\ref{Fig:TMconceptReci}. Because the metasurface transforms the incident and control waves with the same transmission coefficient, by design specification, the power of the incident wave in Fig.~\ref{Fig:TMconceptReci} equally splits into two orthogonally polarized waves. Therefore, using the concept of scattering parameters and denoting the left-hand side of the metasurface as port $1$ and its right-hand side as port $2$, the tensorial backward transmission coefficient $\te{S}_{12}$ reads

\begin{equation}
\label{eq:S12}
\te{S}_{12}= \begin{pmatrix} S_{12}^\text{pp} & S_{12}^\text{ps} \\[5pt] S_{12}^\text{sp} & S_{12}^\text{ss} \end{pmatrix} = \frac{\sqrt{2}}{2}\begin{pmatrix} 0 & -1 \\[5pt] 0 & 1 \end{pmatrix},
\end{equation}
where $S_{12}^\text{pp}$ corresponds to the backward transmission coefficient from parallel to parallel polarization, $S_{12}^\text{ps}$ corresponds to the backward transmission coefficient from perpendicular to parallel polarization, etc. Note that the phase difference $\phi$ can be either present in the control wave or induced by the metasurface, as is here the case (minus sign in $S_{12}^\text{ps}$). It follows, by reciprocity, that the transformation in Fig.~\ref{Fig:TMconcept1} is given by
\begin{equation}
\label{eq:S21}
\te{S}_{21}
=\te{S}_{12}^{\dagger}
=\begin{pmatrix} S_{21}^\text{pp} & S_{21}^\text{ps} \\[5pt] S_{21}^\text{sp} & S_{21}^\text{ss} \end{pmatrix}= \frac{\sqrt{2}}{2}\begin{pmatrix} 0 & 0 \\[5pt] -1 & 1 \end{pmatrix},
\end{equation}

\noindent where $\dagger$ is the transpose operator. This results that the device has a power efficiency that is limited to $50\%$ and that reflection is unavoidable if the metasurface is reciprocal. Similar considerations apply in the oblique incidence case of Fig.~\ref{Fig:TMconcept2}.

\section{Metasurface Design}
\label{Sec:Design}

This section presents the design of the metasurfaces shown in Fig.~\ref{Fig:TMconcept}, using the synthesis technique detailed in~\cite{achouri2014general}. First, the metasurface is described as terms of surface susceptibility tensor continuous functions of space, and it is next discretized into an nonuniform array of scattering particles.

\subsection{Surface Susceptibility Synthesis}
\label{sec:synthesis}

A metasurface, assumed to be subwavelength in thickness, may be modeled as a strictly zero-thickness sheet~\cite{achouri2014general}. Such a sheet may be effectively described as an electromagnetic spatial discontinuity characterized by generalized sheet transition conditions (GSTCs)~\cite{Idemen1973,kuester2003av}. For a metasurface lying in the $x-y$ plane at $z=0$, the GSTCs read
\begin{subequations}
\label{eq:BCfinapp}
\begin{align}
\Delta\ve{H}\times\hat{z}
&=i\omega\ve{P}_{\parallel}+\hat{z}\times\nabla_{\parallel}M_{z},\label{eq:BCfinapp_1}\\
\hat{z}\times\Delta\ve{E}
&=i\omega\mu \ve{M}_{\parallel}+\nabla_{\parallel}\bigg(\frac{P_{z}}{\epsilon }\bigg)\times\hat{z},\label{eq:BCfinapp_2}\\
\hat{z}\cdot\Delta\ve{D}
&=-\nabla\cdot\ve{P}_{\parallel},\\
\hat{z}\cdot\Delta\ve{B}
&=-\mu \nabla\cdot\ve{M}_{\parallel},
\end{align}
\end{subequations}
where $\ve{P}$ and $\ve{M}$ are the electric and magnetic polarization densities, respectively. The operator $\Delta$ denotes the difference of the specified fields between the two sides of the metasurface. The equations~\eqref{eq:BCfinapp_1} and~\eqref{eq:BCfinapp_2} may be transformed into a set of linearly coupled equations if its characterized by purely surface susceptibility elements $(P_z=M_z=0)$. In that case, substituting $\ve{P}$ and $\ve{M}$ by their general bi-anisotropic definition via the susceptibility tensors $\te{\chi}_\text{ee}, \te{\chi}_\text{mm}, \te{\chi}_\text{em}$ and $\te{\chi}_\text{me}$~\cite{lindell1994electromagnetic}, Eqs.~\eqref{eq:BCfinapp} reduce to
\begin{subequations}
\label{eq:BC_plane}
\begin{align}
\Delta\ve{H}\times\hat{z}
&=i\omega\epsilon\te{\chi}_\text{ee}\ve{E}_\text{av}+ik\te{\chi}_\text{em}\ve{H}_\text{av},\label{eq:BC_plane_1}\\
\hat{z}\times\Delta\ve{E}
&=i\omega\mu \te{\chi}_\text{mm}\ve{H}_\text{av}+ik\te{\chi}_\text{me}\ve{E}_\text{av},\label{eq:BC_plane_2}
\end{align}
\end{subequations}
where the polarizations densities have been expressed in terms of the arithmetic averaged of the specified fields at the two sides metasurface~\cite{kuester2003av}.

The system~\eqref{eq:BC_plane} may be exactly solved to provide the metasurface surface susceptibilities for arbitrary specified incident, reflected and transmitted fields~\cite{achouri2014general}. For the transformation in Fig.~\ref{Fig:TMconcept1}, the metasurface is uniform in the $x-y$ plane since all the waves are specified to be normal to the metasurface. In that case, the reflection coefficient tensors have the form
\begin{equation}
\label{eq:S1122}
\te{S}_{11}= \frac{\sqrt{2}}{2}\begin{pmatrix} 0 & e^{i\phi_\text{ps}} \\[5pt] e^{i\phi_\text{sp}} & 0 \end{pmatrix},
\quad
\te{S}_{22}= \begin{pmatrix} e^{i\phi_\text{pp}} & 0 \\[5pt] 0 & 0 \end{pmatrix},
\end{equation}
where $\phi_\text{ab}$ (a,b$=$s,p) are reflection phases corresponding to the specified polarizations, which may be left as free parameters as they are not essential for the required transformations. The metasurface susceptibilities can be straightforwardly found by inserting the electromagnetic fields corresponding to~\eqref{eq:S12},~\eqref{eq:S21} and~\eqref{eq:S1122} into~\eqref{eq:BC_plane}. The corresponding metasurface is fully bi-anisotropic, with $16$ non-zero susceptibility tensor elements, which are omitted here for briefness. The oblique control transformation in Fig.~\ref{Fig:TMconcept2} can be analyzed in a similar fashion, except that the metasurface in that case becomes non-uniform, leading to more complicated susceptibilities tensor functions.

\subsection{Scattering Particle Synthesis}

\begin{figure}[h!]
\begin{center}
\includegraphics[width=0.5\columnwidth]{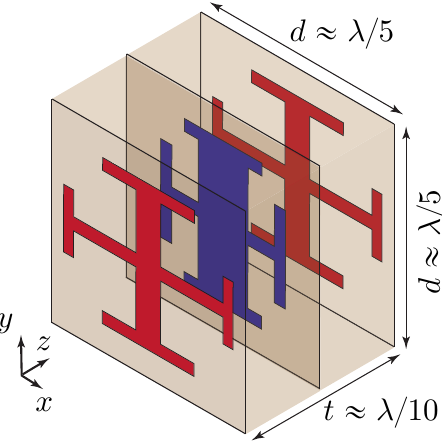}
%\psfragfig*[width=0.5\columnwidth]{cell}{
%\psfrag{a}[c][c][0.9]{$d\approx\lambda/5$}
%\psfrag{b}[c][c][0.9]{$t\approx\lambda/10$}
%\psfrag{x}[c][c][0.9]{$x$}
%\psfrag{y}[c][c][0.9]{$y$}
%\psfrag{z}[c][c][0.9]{$z$}}
\caption{Generic unit cell with three metallic Jerusalem crosses separated by dielectric slabs.}
\label{Fig:metallicUnitCell}
\end{center}
\end{figure}

The next step of the metasurface design consists in discretizing the susceptibility tensor functions and finding the appropriate scattering particles corresponding to each point of the resulting lattice. We use here a scattering particles made of three metallic Jerusalem cross layers held together by two dielectric spacers, as shown in Fig.~\ref{Fig:metallicUnitCell}. Three metallic layers are required to provide full magnitude range ($0-1$) and phase range ($0-2\pi$) transfer functions~\cite{PhysRevApplied.2.044011}, while the Jerusalem cross exhibits relatively well decoupled $x$ and $y$ polarization responses, which simplifies the implementation procedure.

To find the appropriate particle shapes, scattering from the particles is simulated using a commercial software. Coupling between neighboring particles is taken into account using periodic boundary conditions (PBCs) in the $x-y$ plane along with Floquet ports. The simulations yield the tensorial scattering parameters ($\te{S}_{11}, \te{S}_{21}$, etc.) of each scattering particle, which are then related to the original specifications of the metasurface transfer function for determining the particle parameters by iterative analysis~\cite{achouri2014general}.

For instance, the specifications for the uniform, and hence perfectly periodic, metasurface of Fig.~\ref{Fig:TMconcept1} are directly expressed in terms of the scattering parameters~\eqref{eq:S12}, ~\eqref{eq:S21} and~\eqref{eq:S1122}, that may be straightforwardly related to the simulated scattering parameters. In more complicated designs, as that in Fig.~\ref{Fig:TMconcept2}, the susceptibilities are spatially varying in the plane of the metasurface and the initial specifications are given in terms of the electromagnetic fields on the metasurface instead of the scattering parameters, in which case more complex susceptibilitiy-scattering mapping techniques must be used~\cite{achouri2014general,PhysRevApplied.2.044011}.

\section{Full-Wave and Experimental Results}

This section reports the simulation and the experimental results pertaining to the two metasurfaces described in the previous two sections. We first consider the metasurface processor with the normally incident control wave (Fig.~\ref{Fig:TMconcept1}), and next that with the obliquely incident control wave (Fig.~\ref{Fig:TMconcept2}).

\subsection{Normally Incident Control}

\begin{figure}[h!]
\begin{center}
\subfloat[]{\label{Fig:TMstruc1}
\includegraphics[width=0.5\columnwidth]{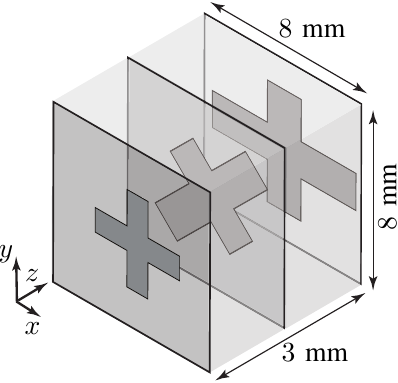}
%\psfragfig*[width=0.45\columnwidth]{struc}{
%\psfrag{a}[c][c][0.8]{$~8~\text{mm}$}
%\psfrag{b}[c][c][0.8]{$~3~\text{mm}$}
%\psfrag{x}[c][c][0.8]{$x$}
%\psfrag{y}[c][c][0.8]{$y$}
%\psfrag{z}[c][c][0.8]{$z$}}
}
\\
\subfloat[]{\label{Fig:TMstruc2}
\includegraphics[width=0.6\columnwidth]{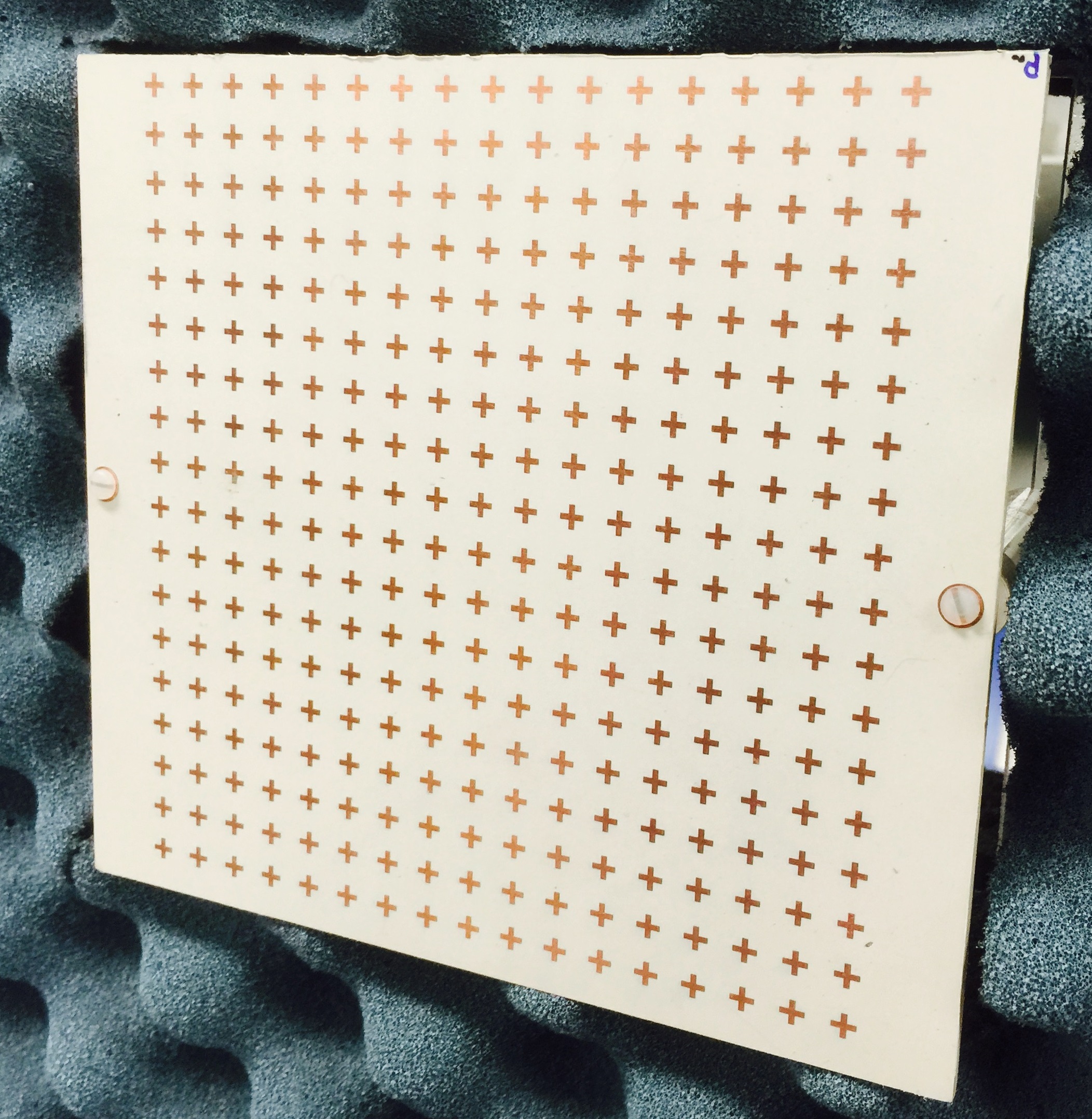}
}
\caption{Switch-amplifier spatial metasurface processor with normally incident control wave~(Fig.~\ref{Fig:TMconcept1}). (a)~Three-layer unit cell designed for an operating frequency of $15$ GHz. (b)~Fabricated structure, composed of $17\times18$ unit cells, in the measurement setup.}\label{Fig:TMstruc11}
\end{center}
\end{figure}

The (periodically repeated) scattering particle required to satisfy the specifications has been implemented following the design procedure outlined in Sec.~\ref{Sec:Design}, and is shown in Fig.~\ref{Fig:TMstruc1}. As seen in this figure, the middle metallic Jerusalem cross is rotated by $45^\circ$ with respect to the outer layer Jerusalem crosses in order to produce the chirality that is required to rotate the polarization of the control wave. The fabricated metasurface is shown in Fig.~\ref{Fig:TMstruc2}.

\begin{figure}[h!]
\centering
\CT
\subfloat[]{\label{Fig:CyN}
\includegraphics[width=0.9\columnwidth]{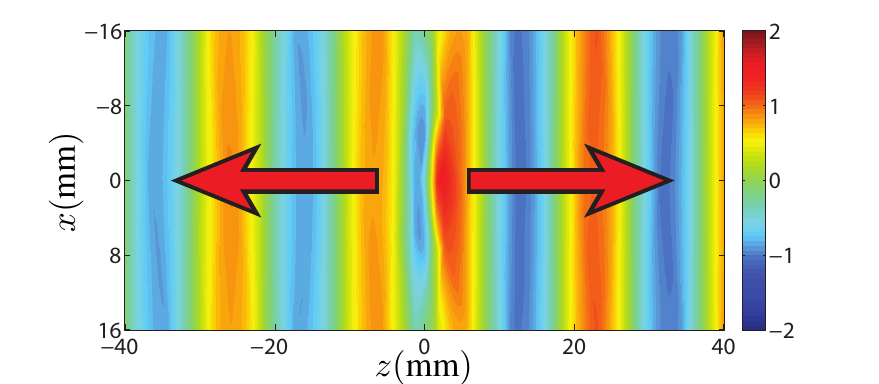}
%\psfragfig*[width=1\columnwidth]{C_y_normal}{
%\psfrag{X}[c][c][1]{$x(\text{mm})$}
%\psfrag{Z}[c][c][1]{$z(\text{mm})$}}
}
\\
\subfloat[]{\label{Fig:SyN}
\includegraphics[width=0.9\columnwidth]{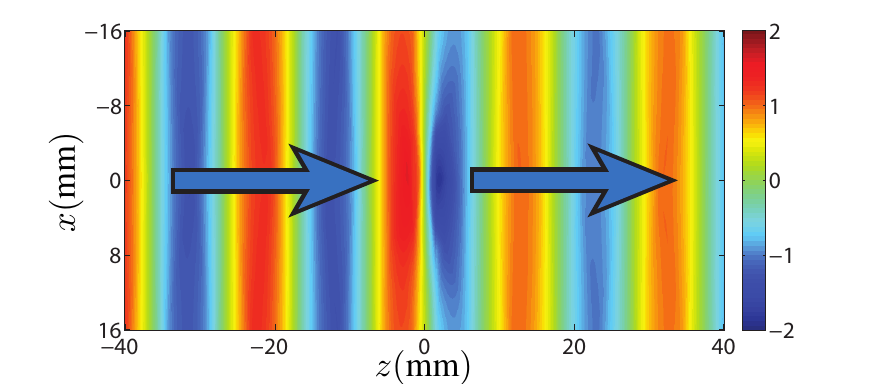}
%\psfragfig*[width=1\columnwidth]{S_y_normal}{
%\psfrag{X}[c][c][1]{$x(\text{mm})$}
%\psfrag{Z}[c][c][1]{$z(\text{mm})$}}
}
\\
\subfloat[]{\label{Fig:CSyN}
\includegraphics[width=0.9\columnwidth]{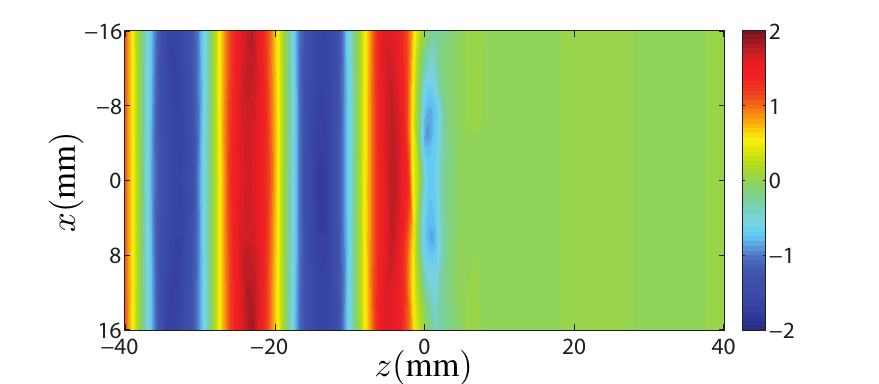}
%\psfragfig*[width=1\columnwidth]{CS_y_normal}{
%\psfrag{X}[c][c][1]{$x(\text{mm})$}
%\psfrag{Z}[c][c][1]{$z(\text{mm})$}}
}
\caption{Full-wave simulated s-polarized fields in the structure in Fig.~\ref{Fig:TMstruc11}. The arrows indicate the direction of wave propagation. The metasurface is located in the plane $z=0$. In (a), only the control wave (initially p-polarized, not shown here) is present, and is partially reflected and transmitted by the metasurface. In (b), only the incident wave is present. In (c), both the incident wave and the control wave are present, and transmission across the metasurface is effectively cancelled by destructive interference.}\label{Fig:TMsimyN}
\end{figure}

\begin{figure}[h!]
\centering
\CT
\subfloat[]{\label{Fig:CxN}
\includegraphics[width=0.9\columnwidth]{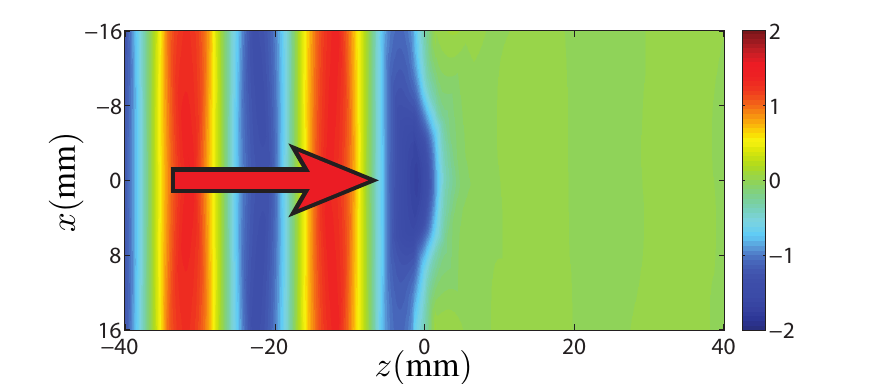}
%\psfragfig*[width=1\columnwidth]{C_x_normal}{
%\psfrag{X}[c][c][1]{$x(\text{mm})$}
%\psfrag{Z}[c][c][1]{$z(\text{mm})$}}
}
\\
\subfloat[]{\label{Fig:SxN}
\includegraphics[width=0.9\columnwidth]{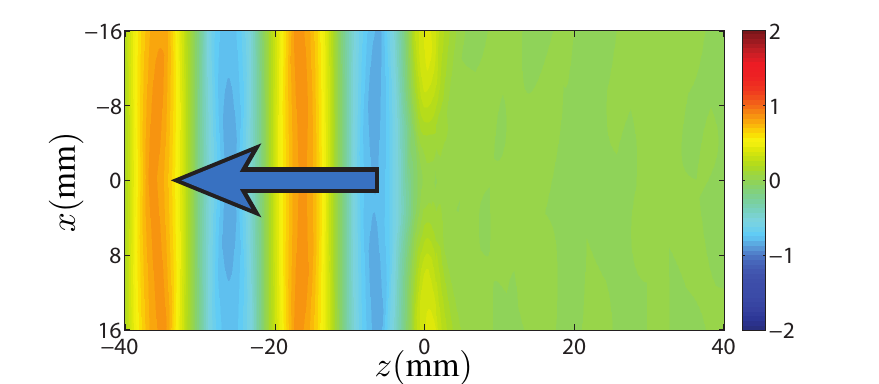}
%\psfragfig*[width=1\columnwidth]{S_x_normal}{
%\psfrag{X}[c][c][1]{$x(\text{mm})$}
%\psfrag{Z}[c][c][1]{$z(\text{mm})$}}
}
\\
\subfloat[]{\label{Fig:CSxN}
\includegraphics[width=0.9\columnwidth]{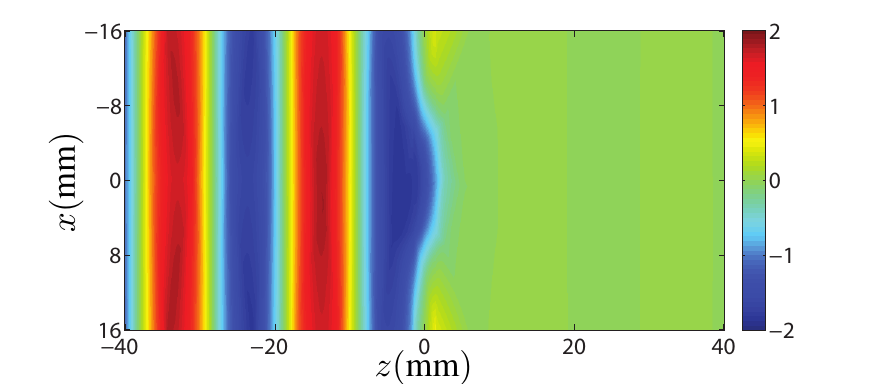}
%\psfragfig*[width=1\columnwidth]{CS_x_normal}{
%\psfrag{X}[c][c][1]{$x(\text{mm})$}
%\psfrag{Z}[c][c][1]{$z(\text{mm})$}}
}
\caption{Same as in Fig.~\ref{Fig:TMsimyN} but for p-polarized fields. In (a), no s-polarized transmission is visible because the metasurface rotates the polarization of the control wave. In (b), the field present on the left-hand side of the metasurface is purely due to the rotated reflection of the incident wave.}\label{Fig:TMsimxN}
\end{figure}

The numerical simulations of the spatial metasurface processor in~Fig.~\ref{Fig:TMstruc} are plotted in Fig.~\ref{Fig:TMsimxN} for the p-polarized waves and in Fig.~\ref{Fig:TMsimyN} for the s-polarized waves. In the Figs.~\ref{Fig:CxN} and~\ref{Fig:CyN} only the control wave is illuminating the metasurface. In both figures, plots (a), (b) and (c) represent the control wave only, the incident wave only, and the superposition of the incident and control waves, respectively, while the arrows indicate the direction of wave propagation.

Figure~\ref{Fig:CxN} shows the control wave (initially p-polarized) that is normally incident onto the metasurface. Since the metasurface rotates the polarization of this wave by $90^\circ$, none of its p-polarized component is transmitted. We see in Fig.~\ref{Fig:CyN} that the control wave is effectively transformed into s-polarization, almost half of it being reflected and the other half being transmitted. Figure~\ref{Fig:SyN} shows the incident wave (initially s-polarized) impinging on the metasurface, with half of it being transmitted without rotation of polarization and the other half being reflected with p-polarization, as shown in Fig.~\ref{Fig:SxN}. One should notice that, the incident and control waves on the right-hand sides of Figs.~\ref{Fig:CyN} and~\ref{Fig:SyN} have both the same amplitude but opposite phases. As a result, as the two waves simultaneously imping on the metasurface, they destructively interfere at the output so as to suppress power transmission, as evidenced in Fig.~\ref{Fig:CSyN}.

\begin{figure}[h!]
\begin{center}
\includegraphics[width=0.9\columnwidth]{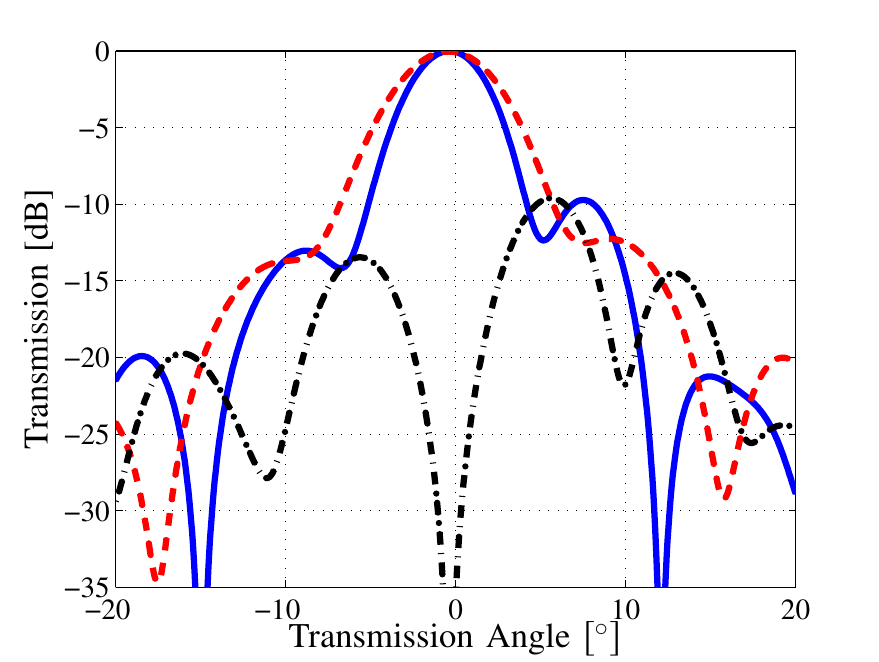}
%\psfragfig*[width=1\columnwidth]{MeasNormal}{
%\psfrag{Y}[c][c][1]{Transmission [dB]}
%\psfrag{X}[c][c][1]{Transmission Angle $[^\circ]$}}
\caption{Experimental results for the metasurface of Fig.~\ref{Fig:TMstruc2} with the metasurface illuminated by the incident wave (continuous blue line), the control wave (dashed red line) and the combination of the two (dashed-dot black line) at $15$~GHz.}
\label{Fig:MeasNormal}
\end{center}
\end{figure}

The experimental results are presented in Fig.~\ref{Fig:MeasNormal}, where the metasurface of Fig.~\ref{Fig:TMstruc2} is illuminated by the incident wave, the control wave and the combination of the two. The spatial metasurface switch exhibits an isolation of over 35~dB. The corresponding spatial metasurface amplification results, obtained by tuning the phase of the control signal for constructive interference at the output, are not shown here for the sake conciseness.

\subsection{Obliquely Incident Control}

\begin{figure}[h!]
\begin{center}
\subfloat[]{\label{Fig:TMstruc11}
\includegraphics[width=0.9\columnwidth]{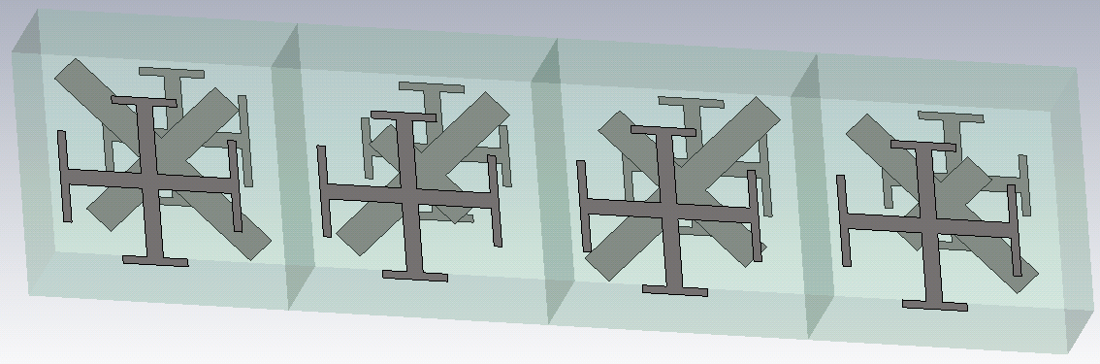}
%\psfragfig*[width=0.45\columnwidth]{struc}{
%\psfrag{a}[c][c][0.8]{$8\text{mm}$}
%\psfrag{b}[c][c][0.8]{$3\text{mm}$}
%\psfrag{x}[c][c][0.8]{$x$}
%\psfrag{y}[c][c][0.8]{$y$}
%\psfrag{z}[c][c][0.8]{$z$}}
}
\\
\subfloat[]{\label{Fig:TMstruc22}
\includegraphics[width=0.6\columnwidth]{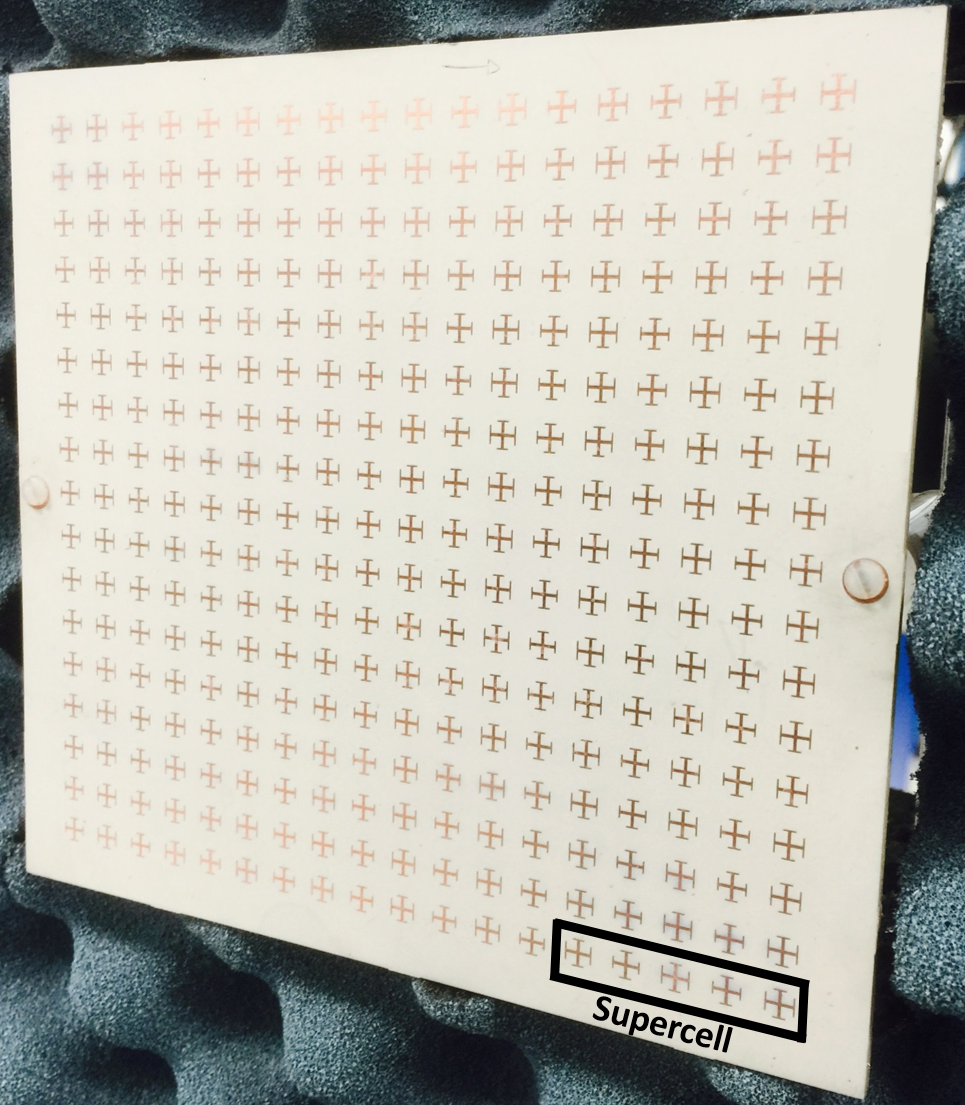}
}
\caption{Switch-amplifier spatial metasurface processor with obliquely incident control wave~(Fig.~\ref{Fig:TMconcept2}). (a) Representation of the supercell composed of four unit cells, the structure is realized for an operating frequency of $10$ GHz. (b) Fabricated structure, with $17\times18$ unit cells, on the measurement stage. The black box indicates where the supercell is on the metasurface and how it is periodically repeated.}\label{Fig:TMstruc222}
\end{center}
\end{figure}

To take into account the metasurface spatial variations, the susceptibility tensor transfer function has been discretized into four distinct unit cells, forming a supercell, as shown in Fig.~\ref{Fig:TMstruc11}. The corresponding fabricated metasurface is shown in Fig.~\ref{Fig:TMstruc22}. From an intuitive perspective, each of the four unit cells induces a different phase shift for the control wave so that this wave gets normally transmitted across the metasurface. The phase shift between adjacent unit cells is $\pi/2$ in order for the four unit cells to cover a complete phase cycle. Note that, due to symmetries, only the two unit cells on the left-hand side of Fig.~\ref{Fig:TMstruc11} were designed, while the two on the right-hand side are their rotated counterparts.

The design procedure requires each unit cell to be simulated in a perfectly periodic and uniform environment. However, the final metasurface is not uniform due to the spatially variant nature of the specified transformation. Therefore, when the unit cells are combined together to realize the final structure, the coupling between them will differ from the original simulations. Consequently, the metasurface response exhibits some undesired scattered fields compared to the expected ideal specifications. These effects may be suppressed with more intensive design efforts.

\begin{figure}[h!]
\centering
\CT
\subfloat[]{\label{Fig:CyO}
\includegraphics[width=0.9\columnwidth]{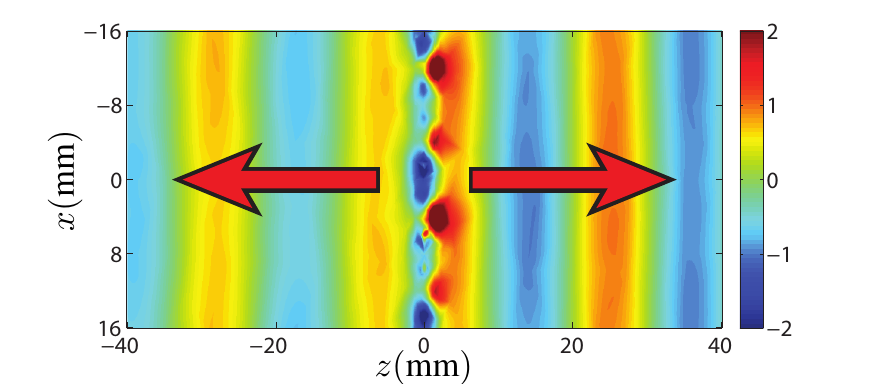}
%\psfragfig*[width=1\columnwidth]{C_y_oblique}{
%\psfrag{X}[c][c][1]{$x(\text{mm})$}
%\psfrag{Z}[c][c][1]{$z(\text{mm})$}}
}
\\
\subfloat[]{\label{Fig:SyO}
\includegraphics[width=0.9\columnwidth]{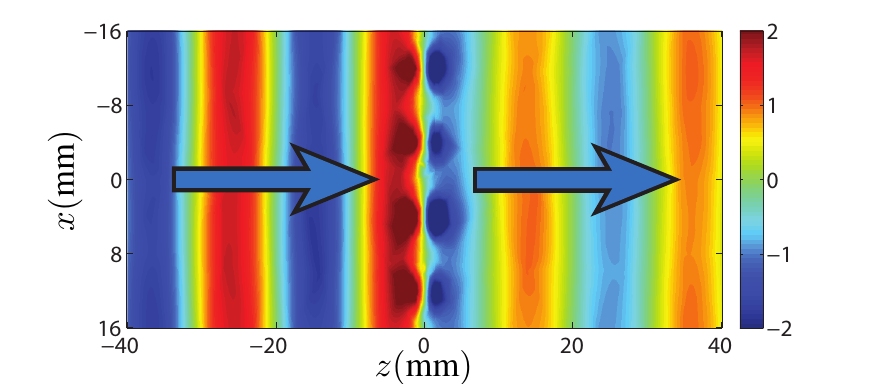}
%\psfragfig*[width=1\columnwidth]{S_y_oblique}{
%\psfrag{X}[c][c][1]{$x(\text{mm})$}
%\psfrag{Z}[c][c][1]{$z(\text{mm})$}}
}
\\
\subfloat[]{\label{Fig:CSyO}
\includegraphics[width=0.9\columnwidth]{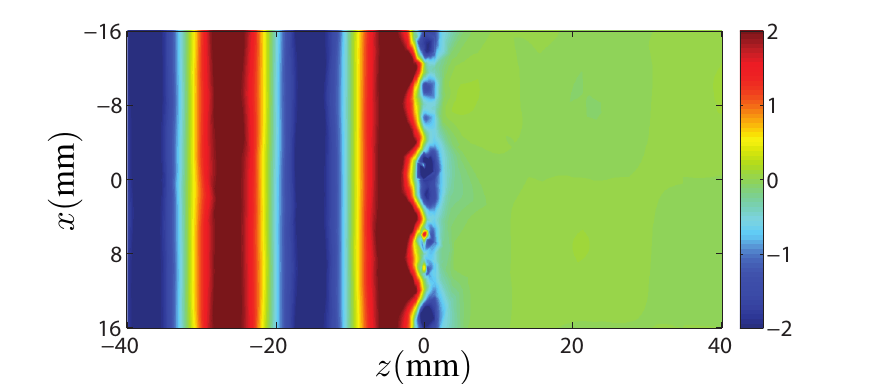}
%\psfragfig*[width=1\columnwidth]{CS_y_oblique}{
%\psfrag{X}[c][c][1]{$x(\text{mm})$}
%\psfrag{Z}[c][c][1]{$z(\text{mm})$}}
}
\caption{Full-wave simulated s-polarized fields in the structure in Fig.~\ref{Fig:TMstruc222}. In (a), only the control wave (initially p-polarized, not shown here) is present, and is partially normally reflected and normally transmitted by the metasurface. In (b), only the incident wave is present. In (c), both the incident wave and the control wave are present, and transmission across the metasurface is effectively cancelled by destructive interference.}\label{Fig:TMsimyO}
\end{figure}

\begin{figure}[h!]
\centering
\CT
\subfloat[]{\label{Fig:CxO}
\includegraphics[width=0.9\columnwidth]{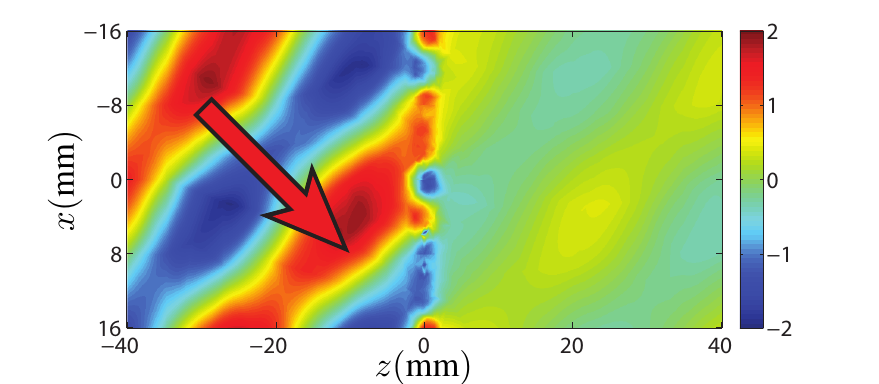}
%\psfragfig*[width=1\columnwidth]{C_x_oblique}{
%\psfrag{X}[c][c][1]{$x(\text{mm})$}
%\psfrag{Z}[c][c][1]{$z(\text{mm})$}}
}
\\
\subfloat[]{\label{Fig:SxO}
\includegraphics[width=0.9\columnwidth]{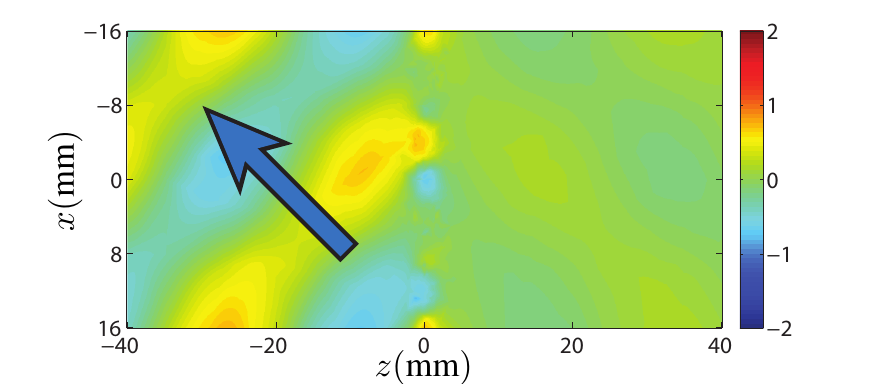}
%\psfragfig*[width=1\columnwidth]{S_x_oblique}{
%\psfrag{X}[c][c][1]{$x(\text{mm})$}
%\psfrag{Z}[c][c][1]{$z(\text{mm})$}}
}
\\
\subfloat[]{\label{Fig:CSxO}
\includegraphics[width=0.9\columnwidth]{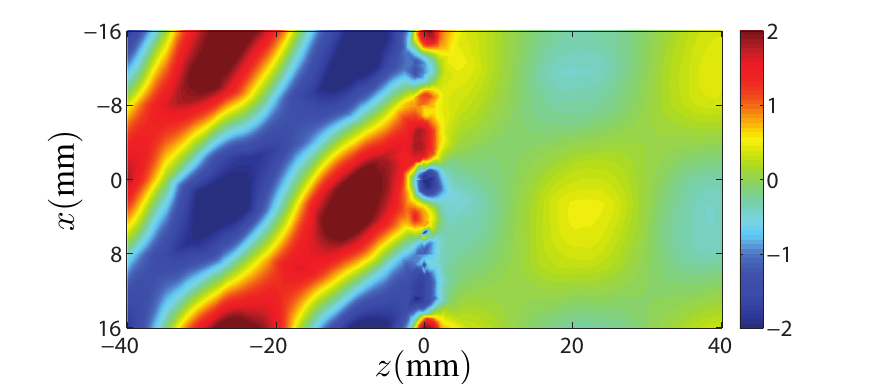}
%\psfragfig*[width=1\columnwidth]{CS_x_oblique}{
%\psfrag{X}[c][c][1]{$x(\text{mm})$}
%\psfrag{Z}[c][c][1]{$z(\text{mm})$}}
}
\caption{Same as in Fig.~\ref{Fig:TMsimyO} but for p-polarized fields. In (a), the control wave is obliquely impinging on the metasurface and no s-polarized transmission is visible because the metasurface rotates the polarization of the wave. In (b), the field present on the left-hand side of the metasurface is due to the rotated reflection of the incident wave.}\label{Fig:TMsimxO}
\end{figure}

The numerical simulations of the metasurface are presented in Fig.~\ref{Fig:TMsimxO} for the p-polarized waves and in Fig.~\ref{Fig:TMsimyO} for the s-polarized waves. The control wave (initially p-polarized) is obliquely incident on the metasurface with a $50^\circ$ angle, as shown in Fig.~\ref{Fig:CxO}. Since the metasurface rotates the polarization of the control wave, almost no p-polarized component is transmitted (at the exception of a small parasitic transmission and reflection). In Fig.~\ref{Fig:CyO}, we see that the control wave is effectively transformed into s-polarization and that almost half of the wave is normally transmitted. In Fig.~\ref{Fig:SyO}, the incident wave (initially s-polarized) is incident on the metasurface and half of it is transmitted without rotation of polarization. The other half of the incident wave is mostly reflected at an angle with p-polarization, as shown in Fig.~\ref{Fig:SxO}. One should notice that, on the right-hand side of Figs.~\ref{Fig:CyO} and~\ref{Fig:SyO}, the incident wave and the control wave have both the same amplitude but opposite phases. When the two waves are simultaneously impinging on the metasurface, the results is a cancellation of all s-polarized transmitted power, as evidenced in Fig.~\ref{Fig:CSyO}.


\begin{figure}[h!]
\begin{center}
\includegraphics[width=0.9\columnwidth]{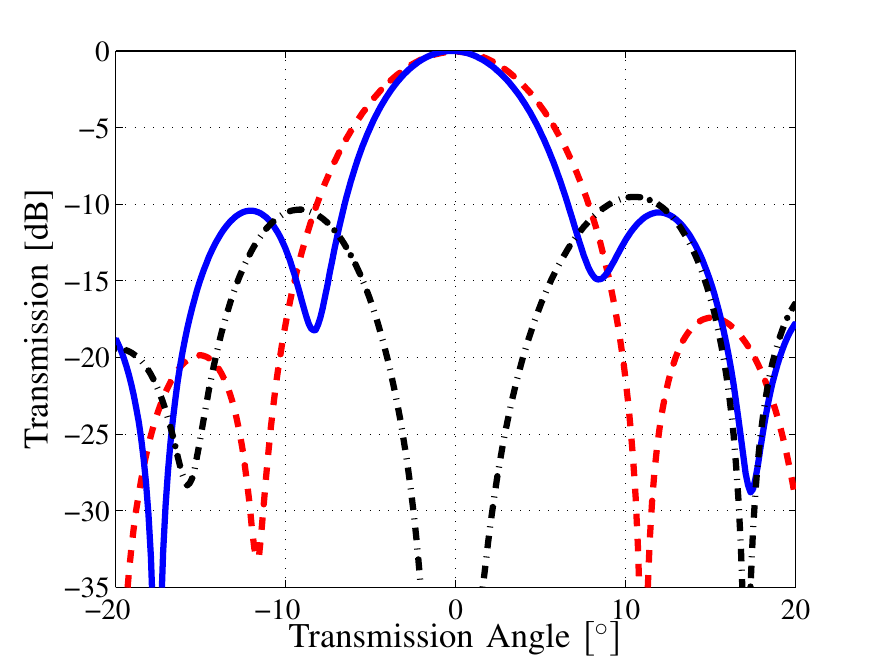}
%\psfragfig*[width=1\columnwidth]{MeasOblique}{
%\psfrag{Y}[c][c][1]{Transmission [dB]}
%\psfrag{X}[c][c][1]{Transmission Angle $[^\circ]$}}
\caption{Experimental results or the metasurface of Fig.~\ref{Fig:TMstruc22} with the metasurface illuminated by the incident wave (continuous blue line), the control wave (dashed red line) and the combination of the two (dashed-dot black line) at $10$~GHz.}
\label{Fig:MeasOblique}
\end{center}
\end{figure}

Finally, the experimental results are presented in Fig.~\ref{Fig:MeasOblique}, where the metasurface of Fig.~\ref{Fig:TMstruc22} is illuminated by the incident wave (smooth line), the control wave (dashed line) and the combination of the two (dot-dashed line). When the two waves are simultaneously present, the cancellation of the transmitted wave is clearly visible.

\section{Conclusion}

We have presented and discussed the concept and implementation of a metasurface spatial processor, whose transmission can be engineered by the application of an electromagnetic control wave. The coherent superposition of the incident wave and the control wave through the metasurface is analogous to the conceptual operation of both a transistor and a Mach-Zehnder interferometer. We have demonstrated two metasurfaces operating as electromagnetic switches or amplifiers. The proposed concept may be extended to a great diversity of electromagnetic modulated wave transformations.

\bibliographystyle{IEEEtran}
\bibliography{LIB}

\end{document}